\documentclass[twocolumn,%preprint,
 showpacs,floats,preprintnumbers,amsmath,amssymb,aps,pre]{revtex4}
\usepackage{epsfig}
\usepackage{graphicx}
\usepackage{amsmath}
\usepackage{CJK}
\begin{document}
\begin{CJK*}{GBK}{song}
\title{Joule-Thomson coefficient of ideal anyons within fractional exclusion statistics}
\author{Fang Qin\footnote{Email: qinfang@phy.ccnu.edu.cn}
and Ji-sheng Chen\footnote{Email: chenjs@iopp.ccnu.edu.cn }}
%\author{Fang Qin (ñû•P)\footnote{Email: qinfang@phy.ccnu.edu.cn}
%and Ji-sheng Chen (³Â¼Ìʤ)\footnote{Email: chenjs@iopp.ccnu.edu.cn}}
\affiliation{Physics Department and Institute of Nanoscience and
Nanotechnology, Central China Normal University, Wuhan 430079,
People's Republic of China}

\begin{abstract}

The analytical expressions of the Joule-Thomson coefficient for
homogeneous and harmonically trapped three-dimensional ideal anyons
which obey Haldane fractional exclusion statistics are derived. For
an ideal Fermi gas, the Joule-Thomson coefficient is negative, which
means that there is no maximum Joule-Thomson inversion temperature.
With careful study, it is found that there exists a Joule-Thomson
inversion temperature in the fractional exclusion statistics model.
Furthermore, the relations between the Joule-Thomson inversion
temperature and the statistical parameter $g$ are investigated.


\end{abstract}


\maketitle
\section{Introduction}

The fractional exclusion statistics of Haldane is an intermediate
statistics between Bose-Einstein and Fermi-Dirac statistics. It is
also a generalized dimensional-independent statistics based on state
counting methods and is suitable to describe interacting
many-particle systems in condensed matter
\cite{Haldane1991,Johnson1994,Wu1994}. By considering $N$ particles
in a $d$-dimensional Hilbert space at fixed size and boundary
conditions, the linear relation between the changes of the
single-particle space $\Delta d$ and the changes of the particle
number $\Delta N$ is defined as $\Delta d=-g\Delta N$ with a
parameter $g$ given by Haldane \cite{Haldane1991}.

In Haldane fractional exclusion statistics, the number of
microscopic quantum states $W$ of $N$ identical particles occupying
a group of $G$ states is \begin{eqnarray}
W=\prod_{i}\frac{[G_{i}+(N_{i}-1)(1-g)]!}{N_{i}![G_{i}-gN_{i}-(1-g)]!},
\end{eqnarray} which is interpolated by Johnson, Canright and Wu \cite{Johnson1994,Wu1994}.
It corresponds to the result of Bose-Einstein statistics when the
weight factor $g=0$ and to the one of Fermi-Dirac statistics when
$g=1$. Here the statistical parameter $g$ is described as the change
in the number of available states when one particle is added to the
system.

The anyon statistical model has been extensively studied in the
literature
\cite{Potter2007,Iguchi1997,Iguchi19972,Su,Sevincli,Nayak1994,Joyce1996,Isakov1996}.
The thermodynamic solution of the one-dimensional ideal anyon gas
which obeys the Haldane statistics is equivalent to the Bethe ansatz
solution of the Calogero-Sutherland model
\cite{BW1995,Bernard1994,Isak1994,Murthy1999}. Furthermore, the
thermodynamic extension to $d\geqslant1$ dimensions of the
Calogero-Sutherland model is explored by Potter et al.
\cite{Potter2007}. The Haldane's fractional exclusion statistics is
widely used to describe the low-dimensional condensed matter
physics. In the literature, the properties of spinons are
characterized in terms of the Haldane's fractional exclusion
statistics with corresponding models ( e.g., the Haldane-Shastry
model in which free spinons exist
\cite{Greiter2005,Greiter2006,Schuricht2008}, the one-dimensional
supersymmetric t-J model in which the thermodynamics of spinons is
investigated \cite{Kuramoto1,Kuramoto2,Sutherland}, and the
Wess-Zumino-Witten model
\cite{Haldane19912,Haldane1992,Bouwknegt1999}). Besides, the Hubbard
model with an infinite-range interaction is also used to study
particles obeying Haldane fractional statistics
\cite{Vitoriano2000,Vitoriano2005}.

In thermodynamics, the fact that the temperature changes with the
decrease of pressure during an adiabatic or throttling expansion is
called the Joule-Thomson effect. The adiabatic expansion or
Joule-Thomson process describes the procedure that a gas is forced
through a porous plug without heat exchange with the environment.
This effect can be described by the Joule-Thomson coefficient
$u_{JT}$, which is the partial derivative of temperature with
respect to pressure at constant enthalpy
\begin{eqnarray}\label{JT} u_{JT}\equiv\left(\frac{\partial T}{\partial
P}\right)_{H},
\end{eqnarray} where $P$ is pressure, $T$ is system temperature and $H$ is enthalpy.
If $u_{JT}>0$, it shows that the temperature of the system decreases
during the adiabatic expansion. On the other hand, if $u_{JT}<0$,
the temperature increases in the throttling process. By setting
$u_{JT}=0$, one can calculate the maximum Joule-Thomson inversion
temperature. Below this inversion temperature, a free adiabatic
expansion causes a decrease in temperature, while it causes a
temperature increase above this inversion temperature
\cite{Pathria1996}.

There have been many works on the Joule-Thomson coefficient. The
Joule-Thomson coefficients of ideal quantum systems are obtained by
Ref. \cite{Saygin2001}. The corresponding results for weakly
interacting Fermi and Bose gases are given by means of the
pseudopotential method \cite{Yuan2007}. Recently, the Joule-Thomson
coefficient for a $d$-dimensional ideal Bose gas was derived in a
power-law potential \cite{Yuan2010}. The Joule-Thomson coefficient
for a strongly interacting Fermi gas was also discussed within the
quasilinear approximation framework \cite{Chen2010}.

According to Landau's phenomenological theory, there is an important
parameter called effective mass which is used to characterize the
quasiparticle and is determined by the low-temperature isochore heat
capacity. If ideal anyons are considered as quasiparticles to
describe interactions \cite{Bhaduri1,Bhaduri2,Bhaduri3,Qin1,Qin2},
the effective mass is a significant physical quantity in the
fractional exclusion statistics model.

In this paper, the analytical expressions of Joule-Thomson
coefficients for homogeneous and harmonically trapped
three-dimensional ideal anyons are derived within the Haldane
fractional exclusion statistics. Furthermore, the Joule-Thomson
inversion temperature varying with the statistical parameter $g$ is
plotted and the effective mass is calculated. The outline is as
follows. In Section \ref{section2}, the distribution function of
Haldane fractional exclusion statistics is obtained by the method of
Lagrange multiplier. The Joule-Thomson coefficients of homogeneous
and harmonically trapped ideal anyons within fractional exclusion
statistics are derived analytically in Section \ref{section3}.
Effective mass is calculated in Section \ref{section4}. The
numerical calculations and the Joule-Thomson inversion temperature
plotted as a function of parameter $g$ are obtained in Section
\ref{section5}. A summary is given in the final section.

\section{Distribution function of Haldane fractional exclusion statistics}\label{section2}

Let us set two Lagrange multipliers $\alpha=-\mu/T$ and $\beta=1/T$,
where $\mu$ is the chemical potential and the natural units
$k_{B}=\hbar=1$ are used. The average occupation number is defined
as $\bar{N_{i}}\equiv N_{i}/G_{i}$. According to the Lagrange
multiplier method $\delta\ln{W}-\alpha\delta N-\beta\delta E=0$, the
most probable distribution function can be derived as \cite{Wu1994}
\begin{eqnarray}\label{f}
\bar{N}=\frac{1}{\omega+g}, \end{eqnarray} where the statistical
parameter $g$ and $\omega$ are related as
\begin{eqnarray}\label{e} \epsilon=\mu+T\left
[(1-g)\ln{(1+\omega)}+g\ln{\omega}\right ],
\end{eqnarray} where $\epsilon$ is the single-particle energy.
When $g=1$, the distribution function $\bar{N}$ goes back to
fermionic distribution function, and when $g=0$ it becomes a bosonic
one. According to equation (\ref{f}), when the temperature is zero,
one can get $\bar{N}=0$ with $\epsilon>\mu$, and $\bar{N}=1/g$ with
$\epsilon<\mu$, which indicates that $g$ characterizes the
generalized Pauli principle since the maximum value of the
occupation number for a single-particle state is $1/g$.

We define the fugacity $z$ in terms of the chemical potential $\mu$
and introduce a parameter $\omega_{0}$ as follows:
\begin{eqnarray}\label{z}
z\equiv exp\left (\frac{\mu}{T}\right
)=(1+\omega_{0})^{g-1}\omega_{0}^{-g}.
\end{eqnarray}

Inserting equation (\ref{z}) into equation (\ref{e}), one obtains
\begin{eqnarray}\label{ee1}
\epsilon=T\left [(1-g)\ln{\left (\frac{1+\omega}{1+\omega_{0}}\right
)}+g\ln{\left (\frac{\omega}{\omega_{0}}\right )}\right ].
\end{eqnarray}

\section{The Joule-Thomson coefficient given by the fractional exclusion statistics}\label{section3}

\subsection{Homogeneous gas}

The density of states is
$D(\epsilon)=(2m)^{3/2}V\epsilon^{1/2}/(2\pi^{2})$ with two degrees
of the spin degeneracy for a homogeneous gas, where $m$ is the
particle mass and $V$ is the system volume.

The expression for the grand thermodynamic potential $\Omega$ is
\cite{Potter2007,Iguchi1997,Iguchi19972,Su,Sevincli}
\begin{eqnarray}\label{P1}
\Omega\nonumber&&=-PV\\
\nonumber&&=-T\int_{0}^{\infty}{D(\epsilon)\ln{\left (1+\frac{1}{\omega}\right )}d\epsilon}\\
&&=-\frac{TV(2m)^{3/2}}{3\pi^{2}}\left [\epsilon^{3/2}\left.
\ln{\left (1+\frac{1}{\omega}\right
)}\right|^{\epsilon=\infty}_{\epsilon=0}\right. \nonumber\\
&& \left. ~~-\int_{\omega_{0}}^{\infty}{\epsilon^{3/2}d\ln{\left
(1+\frac{1}{\omega}\right )}}\right ].
\end{eqnarray}
In the one-dimensional supersymmetric t-J model, three species of
particles will contribute to the grand thermodynamic potential by
applying an external magnetic field, respectively \cite{Kuramoto2}.
In the literature, it is pointed out that the one-dimensional
supersymmetric t-J model is equivalent to Haldane fractional
statistics with properly fixed statistical parameters
\cite{Kuramoto2}. Here, we limit to the three-dimensional free
particles in terms of the Haldane's fractional exclusion statistics
without magnetic field. The corresponding thermodynamic potential is
given by equation (\ref{P1}).

With the help of equation (\ref{ee1}), equation (\ref{P1}) can be
reduced to
\begin{eqnarray}\label{P2}
P=\frac{2T}{\lambda^{3}}G_{5/2}(z,g),
\end{eqnarray}
where the thermal de Broglie wavelength is defined as
$\lambda=\sqrt{2\pi/(mT)}$ and \begin{eqnarray}\label{G1}
G_{n}(z,g)=\frac{1}{\Gamma(n)}\int_{0}^{\infty}{\frac{x^{n-1}dx}{\omega+g}}
\end{eqnarray} is the Calogero-Sutherland integral function
\cite{Potter2007}.
$\Gamma(n)\equiv(n-1)!=\int_{0}^{\infty}{exp(-y)y^{n-1}dy}$ is the
gamma function and $x=\epsilon/T$. The Calogero-Sutherland integral
function has been proved to satisfy the recurrence relation which is
the same as the Bose-Einstein and Fermi-Dirac integral functions
\cite{Potter2007}. The recurrence relation is
\begin{eqnarray}
z\frac{\partial}{\partial z}G_{n}(z,g)=G_{n-1}(z,g).
\end{eqnarray}

By turning the variable $\epsilon$ into $\omega$ through equation
(\ref{ee1}), equation (\ref{G1}) can be reduced to
\begin{eqnarray}\label{G2}
G_{n}(z,g)=\frac{1}{\Gamma(n)}h_{n-1}(\omega_{0},g),
\end{eqnarray} where
\begin{eqnarray}\label{h}
h_{n}(\omega_{0},g)=\int_{\omega_{0}}^{\infty}\frac{\left
\{\ln{\left [\left (\frac{\omega}{\omega_{0}}\right )^{g}\left
(\frac{\omega+1}{\omega_{0}+1}\right )^{1-g}\right ]}\right
\}^{n}d\omega}{\omega(\omega+1)}.
\end{eqnarray}

The finite-temperature particle number density and internal energy
density can be represented as
\begin{eqnarray}\label{n1}
n&&=\frac{1}{V}\int_{0}^{\infty}{\frac{D(\epsilon)d\epsilon}{\omega+g}}\nonumber\\
&&=\frac{2}{\lambda^{3}}G_{3/2}(z,g),\\
\label{E1} \frac{E}{V}&&=\frac{1}{V}\int_{0}^{\infty}{\frac{\epsilon
D(\epsilon)d\epsilon}{\omega+g}}\nonumber\\
&&=\frac{3T}{\lambda^{3}}G_{5/2}(z,g)\nonumber\\
&&=\frac{3}{2}P.
\end{eqnarray}

At zero temperature, the particle number is
$N=(1/g)\int_{0}^{\widetilde{E}_{F}}{D(\epsilon)d\epsilon}=(2mE_{F})^{3/2}V/(3\pi^{2})$,
where $E_{F}$ is the uniform ideal Fermi energy and
$\widetilde{E}_{F}$ satisfies $\widetilde{E}_{F}=g^{{2}/{3}}E_{F}$.
By replacing the particle number of the ground state into equation
(\ref{n1}), one can get
\begin{eqnarray}\label{n2}
\frac{3\pi^{1/2}}{4}\left (\frac{T}{T_{F}}\right
)^{3/2}G_{3/2}(z,g)\nonumber&&=\frac{3}{2}\left
(\frac{T}{T_{F}}\right
)^{3/2}h_{1/2}(\omega_{0},g)\nonumber\\
&&=1
\end{eqnarray} with the Fermi characteristic temperature $T_{F}$ for a uniform ideal Fermi gas.

From equation (\ref{n1}), the partial derivative of particle number
density $n$ with respect to temperature $T$ at constant $n$ is
written as
\begin{eqnarray}\label{*2} \left (\frac{\partial n}{\partial
T}\right )_{n}&&=2\left (\frac{m}{2\pi}\right )^{3/2}T^{1/2}\left
[\frac{3}{2}G_{3/2}(z,g)\right. \nonumber\\
&& \left. ~~+T\left (\frac{\partial
G_{3/2}(z,g)}{\partial T}\right )_{\mu}\right. \nonumber\\
&& \left. ~~+T\left (\frac{\partial G_{3/2}(z,g)}{\partial
\mu}\right )_{T}\left (\frac{\partial \mu}{\partial T}\right
)_{n}\right ].
\end{eqnarray} By combining equation (\ref{*2}) with $(\partial n/\partial
T)_{n}=0$, the partial derivative of chemical potential to
temperature takes the form
\begin{eqnarray}\label{mun}
\left (\frac{\partial \mu}{\partial T}\right
)_{n}=\ln{z}-\frac{3G_{3/2}(z,g)}{2G_{1/2}(z,g)}.
\end{eqnarray}

According to the constant total particle number $N$, one can have
\begin{eqnarray}\label{uniform12}
\left (\frac{\partial P}{\partial T}\right )_{V}\nonumber&&=\left
(\frac{\partial P}{\partial T}\right )_{N,V}\\
&&=\frac{1}{\lambda^{3}}\left
[5G_{5/2}(z,g)-\frac{3G_{3/2}^{2}(z,g)}{G_{1/2}(z,g)}\right ],
\\\label{uniform22} \left (\frac{\partial V}{\partial T}\right
)_{P}\nonumber&&=N\left (\frac{\partial (1/n)}{\partial T}\right
)_{P}\\&&=\frac{N\lambda^{3}}{4TG_{3/2}(z,g)}\nonumber\\
&&~~\times\left
[\frac{5G_{5/2}(z,g)G_{1/2}(z,g)}{G^{2}_{3/2}(z,g)}-3\right ].
\end{eqnarray}

To derive the expression of the Joule-Thomson coefficient, the
isochore heat capacity $C_{V}$ and isobar heat capacity $C_{P}$ per
particle are derived first as
\begin{eqnarray}
\frac{C_{V}}{N}&&\nonumber=\frac{1}{N}\left (\frac{\partial E}{\partial T}\right )_{N,V}\\
&&\label{uniformCV1}=\frac{15G_{5/2}(z,g)}{4G_{3/2}(z,g)}-\frac{9G_{3/2}(z,g)}{4G_{1/2}(z,g)}\\
&&\label{uniformCV2}=\frac{5h_{3/2}(\omega_{0},g)}{2h_{1/2}(\omega_{0},g)}-\frac{9h_{1/2}(\omega_{0},g)}{2h_{-1/2}(\omega_{0},g)},\\
\label{uniformCP}
\frac{C_{P}}{N}&&\nonumber=\frac{C_{V}}{N}+\frac{T}{N}\left (\frac{\partial P}{\partial T}\right )_{V}\left (\frac{\partial V}{\partial T}\right )_{P}\\
&&=\frac{5h_{3/2}(\omega_{0},g)}{6h_{1/2}(\omega_{0},g)}\nonumber\\
&&~~\times\left[\frac{5h_{3/2}(\omega_{0},g)h_{-1/2}(\omega_{0},g)}{3h_{1/2}^{2}(\omega_{0},g)}-3\right].
\end{eqnarray}

Furthermore, from the fundamental thermodynamic relations and
equations (\ref{JT}), (\ref{uniform22}) and (\ref{uniformCP}), the
Joule-Thomson coefficient can be given by
\begin{eqnarray}\label{uniformJT}
u_{JT}&&\nonumber=\frac{1}{C_{P}}\left [T\left (\frac{\partial
V}{\partial T}\right )_{P}-V\right ]\\
&&=\frac{\pi^{1/2}\lambda^{3}}{2}\left[\frac{1}{2h_{3/2}(\omega_{0},g)}\right. \nonumber\\
&& \left.
~~-\frac{h_{-1/2}(\omega_{0},g)}{5h_{3/2}(\omega_{0},g)h_{-1/2}(\omega_{0},g)-9h_{1/2}^{2}(\omega_{0},g)}\right]. \nonumber\\
\end{eqnarray}

\subsection{Harmonically trapped gas}

We define the geometric mean of the trap frequencies as
$\varpi=(\omega_{x}\omega_{y}\omega_{z})^{1/3}$. The corresponding
density of states is $D(\epsilon)=\epsilon^{2}/\varpi^{3}$. In the
similar way as described in the homogeneous case, the Joule-Thomson
coefficient of a trapped gas is given as
\begin{eqnarray}\label{trappedJT} u_{JT}&&\nonumber=\frac{V}{6}\left
(\frac{\varpi}{T}\right )^{3}\left
[\frac{1}{G_{4}(z,g)}\right. \\
\nonumber&& \left.
~~-\frac{G_{2}(z,g)}{4G_{4}(z,g)G_{2}(z,g)-3G_{3}^{2}(z,g)}\right
]\\
&&=\frac{1}{T^{3}}\left
[\frac{1}{h_{3}(\omega_{0},g)}\right. \nonumber\\
&& \left.
~~-\frac{2h_{1}(\omega_{0},g)}{8h_{3}(\omega_{0},g)h_{1}(\omega_{0},g)-9h_{2}^{2}(\omega_{0},g)}\right
],
\end{eqnarray} with $V=\varpi^{-3}$ \cite{Sevincli,Romero}.

The corresponding analytical expressions for the particle number and
isobar heat capacity per particle are
\begin{eqnarray}\label{N1} N=2\left (\frac{T}{\varpi}\right
)^{3}G_{3}(z,g),\end{eqnarray}
\begin{eqnarray}\frac{C_{P}}{N}=\frac{4h_{3}(\omega_{0},g)}{3h_{2}(\omega_{0},g)}\left
[\frac{8h_{3}(\omega_{0},g)h_{1}(\omega_{0},g)}{3h_{2}^{2}(\omega_{0},g)}-3\right
].
\end{eqnarray}

At zero temperature, the particle number is
\begin{eqnarray}\label{N0}
N=\frac{1}{3}\left (\frac{E_{F}}{\varpi}\right )^{3}.
\end{eqnarray}

Substituting equation (\ref{N0}) into equation (\ref{N1}), one gets
\begin{eqnarray}\label{N2}6\left (\frac{T}{T_{F}}\right )^{3}G_{3}(z,g)
\nonumber&&=3\left(\frac{T}{T_{F}}\right )^{3}h_{2}(\omega_{0},g)\nonumber\\
&&=1.
\end{eqnarray}

\section{Effective mass}\label{section4}

At very low temperature, the $G_{n}(z,g)$ can be expanded as
\cite{Potter2007,Nayak1994,Joyce1996,Isakov1996}
\begin{eqnarray}\label{LG}
G_{n}(z,g)=\frac{(\ln{z})^{n}}{g\Gamma(n+1)}\left
[1+\frac{\pi^2}{6}\frac{gn(n-1)}{(\ln{z})^{2}}+\cdot\cdot\cdot\right
].
\end{eqnarray}

From equations (\ref{uniformCV1}) and (\ref{LG}), the expression of
the isochore heat capacity at extremely low temperature can be given
by
\begin{eqnarray}\label{LCV}
\frac{C_{V}}{N}=\frac{\pi^{2}}{2}g^{1/3}\left (\frac{T}{T_{F}}\right
)+\cdot\cdot\cdot.
\end{eqnarray}

The zero-temperature effective mass is
\begin{eqnarray}\label{m}
\frac{m^{*}}{m}&&\nonumber=\frac{C_{V}}{(C_{V})_{ideal}}\\&&=g^{1/3},
\end{eqnarray} where $(C_{V})_{ideal}=\pi^{2}TN/(2T_{F})$ is the isochore heat
capacity of an ideal Fermi gas at the first-order approximation and
$m$ is the mass of non-interacting ideal fermions.

\section{Discussions}\label{section5}

The numerical results will be given in this section.

The Joule-Thomson coefficient plotted as a function of reduced
temperature can be obtained from equations (\ref{n2}) and
(\ref{uniformJT}) for a homogeneous gas of anyons. It can also be
given by equations (\ref{trappedJT}) and (\ref{N2}) for a trapped
anyon gas numerically.

\begin{figure}[htb]
  \centering
   \includegraphics[width = .45\textwidth]{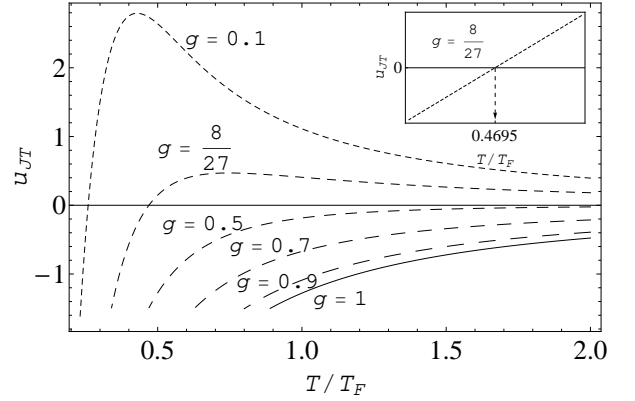}
   \caption{\small
    The Joule-Thomson coefficient $u_{JT}$ is plotted as a
    function of the reduced temperature for homogeneous gases.
    The solid curve denotes that for the ideal fermions,
    and the dashed curves denote the ones for ideal
    anyons with different statistical parameter $g=0.1,\frac{8}{27},0.5,0.7,0.9$.
    Inset shows that the $u_{JT}$
    with fixed $g=\frac{8}{27}$ changes its sign at $T/T_{F}\approx0.4695$.
    $m=1$ and $E_{F}=1$ are chosen for convenience.
    }\label{fig1}
\end{figure}

\begin{figure}[htb]
  \centering
   \includegraphics[width = .45\textwidth]{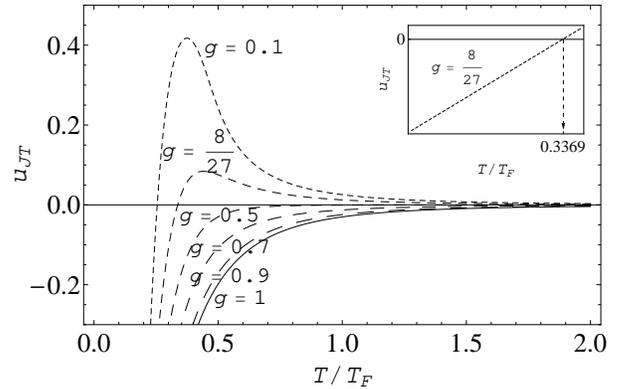}
   \caption{\small
    The line styles are similar to the ones in figure \ref{fig1}
    for harmonically trapped gases. The $u_{JT}$
    with $g=\frac{8}{27}$ changes its sign at $T/T_{F}\approx0.3369$ in the inset.
    Here, $E_{F}=1$.
    }\label{fig2}
\end{figure}

Figures \ref{fig1} and \ref{fig2} show that the Joule-Thomson
coefficient is larger for smaller values of $g$ at the same
temperature for both homogeneous and trapped anyons. Besides, the
curves all tend to zero in the high temperature limit as a Boltzmann
gas and approach the minus infinity in the strong degenerate limit
$T\rightarrow0$. The Joule-Thomson coefficients for both homogeneous
and trapped ideal Fermi gases are always negative, which means that
there is no Joule-Thomson inversion temperature for ideal Fermi gas.
During the adiabatic expansion, the temperature always increases for
an ideal Fermi gas. For ideal anyons, however, there exists a
Joule-Thomson inversion temperature, which depends on the parameter
$g$ in the fractional exclusion statistics model.

Recently, the strong interaction in two-component ultracold fermions
is a hot topic
\cite{Giorgini2008,Bloch2008,Ho2004,Ho20042,Xiong,Qijin}. The
strongly interacting Fermi gas is called the unitary Fermi gas
\cite{Ho2004,Ho20042,Xiong,Qijin}. As a hypothesis, the
three-dimensional ideal anyons with fractional exclusion statistics
can be used to model the statistical behavior of a real unitary
Fermi system \cite{Bhaduri1,Bhaduri2,Bhaduri3,Qin1,Qin2}. This
fractional exclusion statistics hypothesis is found to be in good
agreement with the experimental results in a harmonic trap
\cite{Qin2}.

The parameter $g$ in the statistical model can be fixed as
$g=\xi^{3/2}=\frac{8}{27}$ \cite{Qin1,Qin2}, where the universal
constant $\xi=\frac{4}{9}$ is given by the developed quasi-linear
approximation
\cite{Chenjs2007,Chenjs2008,Chenjs20082,Chenjs2009,chen20092,Chenjs20102}.
By taking $g=\frac{8}{27}$ as an example, the inversion temperature
of a homogeneous unitary Fermi gas is $T/T_{F}\approx0.4695$, and it
is $T/T_{F}\approx0.3369$ for a harmonically trapped unitary system
as indicated in figures \ref{fig1} and \ref{fig2}.

\begin{figure}[htb]
  \centering
   \includegraphics[width = .45\textwidth]{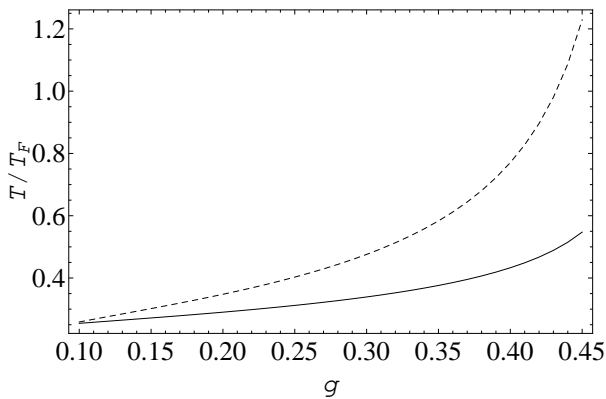}
   \caption{\small
    The reduced Joule-Thomson inversion
    temperature $T/T_{F}$ versus $g$.
    The dashed curve denotes the result of homogeneous anyon gas,
    and the solid curve represents that of trapped gas.
    }\label{fig3}
\end{figure}

It is evident from figure \ref{fig3} that the reduced Joule-Thomson
inversion temperature increases with the increasing of the
statistical parameter $g$. Further, the reduced inversion
temperature of a homogeneous gas is higher than the corresponding
one of a trapped gas with the same value of $g$. The difference in
the reduced inversion temperatures between a homogeneous gas and a
trapped gas becomes larger as $g$ increases.

\section{Summary}\label{section6}

The Joule-Thomson coefficients of a homogeneous and a trapped
three-dimensional ideal anyon system have been analyzed within the
Haldane fractional exclusion statistics. The results show that the
Joule-Thomson coefficient of anyons will overlap with that of the
ideal Fermi gas in the high-temperature Boltzmann regime and
approach the minus infinity in the strong degenerate limit whatever
value of $g$ is chosen. The Joule-Thomson coefficient gets larger
for smaller values of the statistical parameter $g$ at the same
temperature for both ideal homogeneous and trapped anyons. For ideal
fermions, the Joule-Thomson coefficient is negative, which means
that there is no Joule-Thomson inversion temperature. However, there
exists an inversion temperature for the system obeying Haldane
fractional exclusion statistics. The value of the Joule-Thomson
inversion temperature depends on the statistical parameter $g$ in
the fractional exclusion statistics, and it increases with the
increasing $g$. Besides, the reduced inversion temperature of a
homogeneous anyon gas is higher than the corresponding trapped one
with the same $g$. The deviation between the reduced inversion
temperature of a homogeneous gas and that of a trapped gas will
become larger and larger as $g$ increases. The effective mass is
$m^{*}/m=g^{1/3}$, which is a function of the statistical factor
$g$.

\acknowledgments{This work was supported by the National Natural
Science Foundation of China under Grant No. 10875050 and
self-determined research funds of CCNU from the colleges' basic
research and operation of MOE.}

\end{CJK*}
\end{document}